\documentclass[
  reprint,
  superscriptaddress,
  amsmath,amssymb,
  aps,longbibliography,
  floatfix,
]{revtex4-1}

\usepackage{graphicx}
\usepackage{dcolumn}
\usepackage{bm}
\usepackage{amsmath,amssymb,amstext,amsthm}
\usepackage{braket}
\usepackage{bbold}
\usepackage{bbm}
\usepackage{xcolor}
\usepackage{ gensymb }
\usepackage{soul}
\setstcolor{red}

\begin{document}

\setlength{\abovedisplayskip}{1pt}
\title{Topological Expansion of Boehm’s Brushes via Structured Light}

\author{D. A. Pushin}
\altaffiliation{These authors contributed equally to this work.}
\affiliation{Institute for Quantum Computing, University of Waterloo,  Waterloo, ON, Canada, N2L3G1}
\affiliation{Department of Physics, University of Waterloo, Waterloo, ON, Canada, N2L3G1}
\affiliation{Centre for Eye and Vision Research, 17W Hong Kong Science Park, Hong Kong}
\affiliation{Incoherent Vision Inc., Wellesley, ON, Canada, N0B2T0}
\affiliation{School of Optometry and Vision Science, University of Waterloo, Waterloo, ON, Canada, N2L3G1}

\author{I. Salehi} 
\altaffiliation{These authors contributed equally to this work.}

\affiliation{Institute for Quantum Computing, University of Waterloo,  Waterloo, ON, Canada, N2L3G1}
\affiliation{Centre for Eye and Vision Research, 17W Hong Kong Science Park, Hong Kong}
\affiliation{School of Optometry and Vision Science, University of Waterloo, Waterloo, ON, Canada, N2L3G1}

\author{A. Chow} 

\affiliation{Centre for Eye and Vision Research, 17W Hong Kong Science Park, Hong Kong}
\affiliation{School of Optometry and Vision Science, University of Waterloo, Waterloo, ON, Canada, N2L3G1}

\author{A. E. Silva} 
\affiliation{School of Optometry and Vision Science, University of Waterloo, Waterloo, ON, Canada, N2L3G1}
\affiliation{Department of Psychology, Idaho State University, Pocatello, ID, 83209, USA}

\author{P. Chahal}
\affiliation{Department of Physics, University at Buffalo, State University of New York, Buffalo, New York 14260, USA}

\author{D. G. Cory}
\affiliation{Institute for Quantum Computing, University of Waterloo,  Waterloo, ON, Canada, N2L3G1}
\affiliation{Department of Chemistry, University of Waterloo, Waterloo, ON, Canada, N2L3G1}
\author{M. Kulmaganbetov}
\affiliation{Centre for Eye and Vision Research, 17W Hong Kong Science Park, Hong Kong}

\author{G. P. Misson} 
\affiliation{School of Optometry, Aston University, Birmingham B4 7ET, UK}

\author{N. Shentevski}

\affiliation{Department of Physics, University at Buffalo, State University of New York, Buffalo, New York 14260, USA}

\author{T. Singh}
\affiliation{Centre for Eye and Vision Research, 17W Hong Kong Science Park, Hong Kong}
\author{S. E. Temple} 
\affiliation{School of Optometry, Aston University, Birmingham B4 7ET, UK}
\affiliation{Division of Research and Innovation, University of Bristol, Bristol, BS8 1QU, UK}
\affiliation{Azul Optics Ltd, Henleaze, Bristol BS9 4QG, UK}
\author{B. Thompson}
\affiliation{Centre for Eye and Vision Research, 17W Hong Kong Science Park, Hong Kong}
\affiliation{School of Optometry and Vision Science, University of Waterloo, Waterloo, ON, Canada, N2L3G1}
\author{D. Sarenac}
\email{dusansar@buffalo.edu}

\affiliation{Centre for Eye and Vision Research, 17W Hong Kong Science Park, Hong Kong}
\affiliation{Incoherent Vision Inc., Wellesley, ON, Canada, N0B2T0}
\affiliation{School of Optometry and Vision Science, University of Waterloo, Waterloo, ON, Canada, N2L3G1}
\affiliation{Department of Physics, University at Buffalo, State University of New York, Buffalo, New York 14260, USA}

\date{\today}




\begin{abstract}
We report a novel entoptic phenomenon in which the classical two-lobed Boehm’s brushes are transformed into a multi-lobed structure by projecting spin–orbit coupled light onto the human retina. These structured beams, composed of non-separable superpositions of circular polarization and orbital angular momentum (OAM), produce azimuthally modulated entoptic patterns through polarization-dependent scattering in the retina. Unlike Haidinger’s brushes, which arise from dichroic absorption in the macula, the observed effect is driven by angular variations in scattering strength relative to the local polarization direction. In regions where scattering centers exhibit polarization orientations that converge toward a common point, their combined contributions reinforce one another, producing brighter and more sharply defined entoptic lobes whose number and orientation vary systematically with the topology of the spin–orbit stimulus. Psychophysical measurements across retinal eccentricities from 0.5$^\circ$ to 4$^\circ$ in eleven participants revealed that contrast detection thresholds decreased exponentially with eccentricity, consistent with polarization-sensitive scattering by isotropic structures in the non-foveal retinal regions. From the psychophysical fits, the mean eccentricity at which the entoptic pattern reached a 50\% threshold was $r_{50} = 1.03^\circ$ with a 95\% confidence interval of [0.72, 1.34]$^\circ$, indicating that the spin–orbit–induced entoptic structure becomes perceptually robust at approximately 1$^\circ$ retinal eccentricity.  Together, these findings demonstrate that spin–orbit light modulates scattering-based visual phenomena in previously unrecognized ways, enabling new approaches for probing retinal structure and visual processing using topological features of light.
\end{abstract}

\maketitle

\section{Introduction}

Structured light—optical fields with tailored amplitude, phase, and polarization profiles—has opened new frontiers in imaging, communication, and quantum information~\cite{ rubinsztein2016roadmap,bliokh2023roadmap,chen2021engineering,ni2021multidimensional}. Among the most versatile classes of structured beams are those carrying orbital angular momentum (OAM), whose helical wavefronts enable control over light’s spatial structure at both classical and quantum levels~\cite{allen1992orbital, shen2019optical,mair2001entanglement,Andersen2006,cameron2021remote,ritsch2017orbital, wang2012terabit}. When OAM is coupled to polarization, the resulting spin–orbit beams exhibit space-varying polarization topologies and non-separable vectorial modes that can encode rich geometric and topological information~\cite{maurer2007tailoring,sarenac2018generation,marrucci2006optical}.

The human visual system, though largely insensitive to polarization under normal viewing conditions, is capable of perceiving subtle polarization-induced entoptic phenomena~\cite{horvath2004polarized,o2021seeing,temple2024human}. The best-known example is Haidinger’s brushes are a faint, hourglass-shaped pattern visible in central vision when viewing linearly polarized light that includes blue wavelengths~\cite{haidinger1844ueber,misson2020polarization,mottes2022haidinger}. It is thought to arise from dichroic absorption by the aligned macular pigment molecules in a radially symmetric Henle fiber layer~\cite{misson2015human,misson2017spectral,misson2019computational,wang2022mathematical}. Another lesser-known but distinct phenomenon is Boehm’s brushes~\cite{boehm1940neues,o2021seeing}, which appear as a bowtie-shaped pattern in peripheral vision when a small, linearly polarized point source is viewed in the periphery. Unlike Haidinger’s brushes, which arises from dichroic absorption in the macula, Boehm’s brushes is attributed to polarization-sensitive scattering from subcellular structures within the inner retina, particularly in layers such as the inner plexiform layer and ganglion cell layer~\cite{horvath2004polarized,vos1964contribution,weale1976spectral,o2021seeing}.

\begin{figure*}
    \centering
    \includegraphics[width=\linewidth]{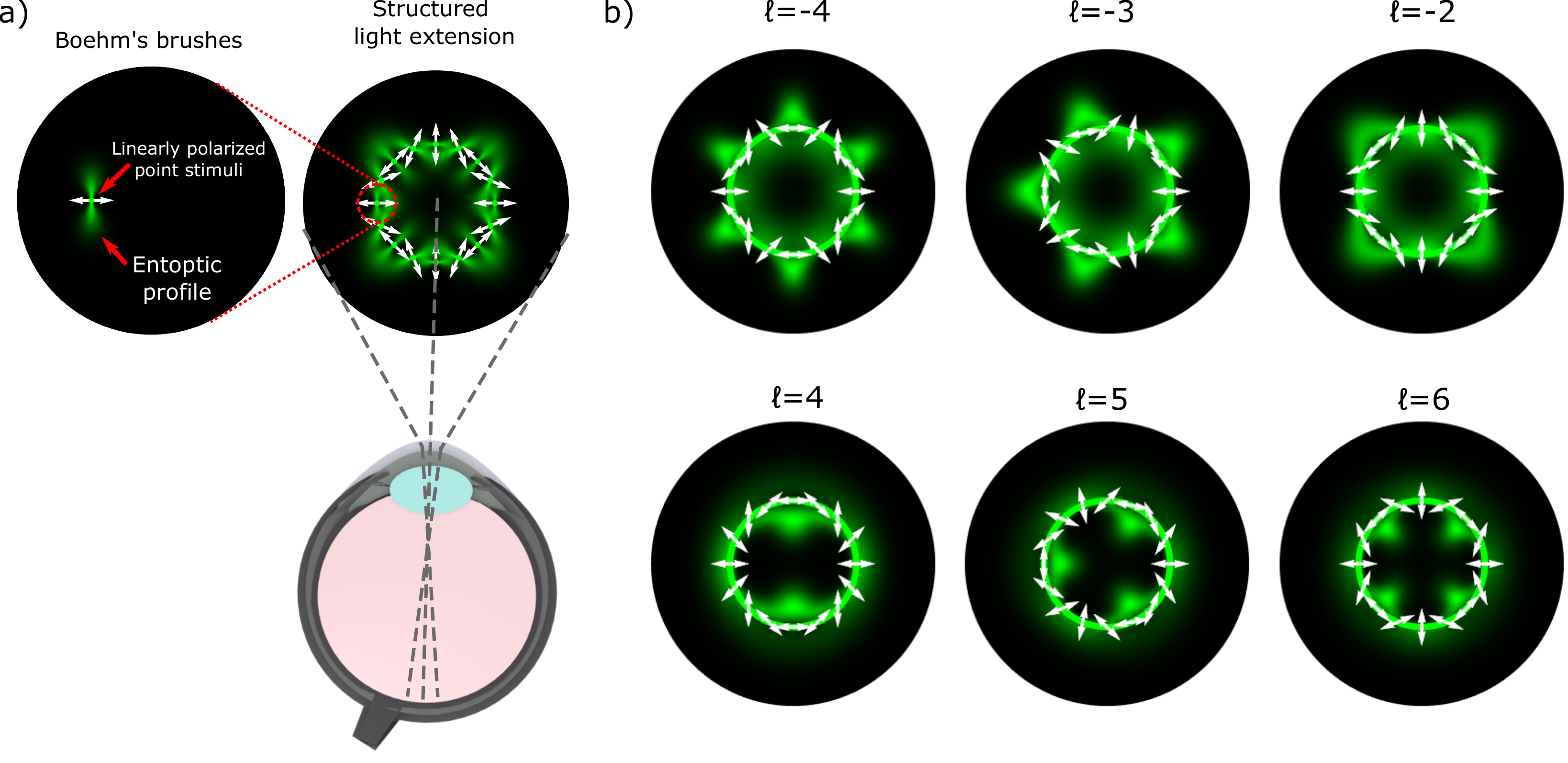}
    \caption{a) Conceptual illustration of how a structured entoptic pattern emerges from spin–orbit light carrying orbital angular momentum (OAM) $\ell = -2$. For illustrative clarity, the stimulus is shown here as a discrete ring of points, each with a polarization orientation determined by the local structure of the spin–orbit state, see Eq.~\ref{Eqn:PsiInGeneral}. Each point elicits Boehm’s brushes through polarization-sensitive scattering, and the perceived global pattern arises from the incoherent sum of these local responses. b) Simulated entoptic profiles corresponding to structured beams with OAM ranging from $\ell = -4$ to $\ell = +6$, viewed from an annular (ring-shaped) stimulus. The number of visible lobes follows $N = |\ell - 2|$. For $\ell < 2$, the polarization modulation is pronounced outside the aperture while for $\ell > 2$ the modulation shifts inward, and the structured pattern appears within the dark central region. The annular aperture profile allows access to the inner polarization-dependent scattering response that is otherwise masked by the high central intensity of a disk stimulus. 
}
    \label{fig:fig1}
\end{figure*}

Classical entoptic patterns have historically been limited to uniform or linearly polarized fields. Recent work has shown that spin–orbit structured light can produce entoptic patterns with azimuthal lobes whose geometry encodes the light’s polarization topology~\cite{sarenac2020direct,sarenac2022human,kapahi2024measuring,pushin2024psychophysical}. These stimuli have been used to modulate Haidinger’s brushes and quantify perceptual thresholds in the macula, enabling discrimination of OAM states and enhancing visibility through spatially varying polarization~\cite{pushin2023structured}. More recently, structured light has been applied to selectively characterize circularly oriented macular pigment~\cite{pushin2025characterizing}, yielding eccentricity-dependent models of optical density based on threshold detection of rotating entoptic patterns. However, these effects have thus far been limited to central, absorption-based mechanisms. Whether polarization topology can also drive entoptic phenomena through peripheral scattering remained unknown, motivating the present study of Boehm’s brushes under structured light illumination.

In this work, we report the discovery of a novel entoptic phenomenon in which the classical two-lobed Boehm’s brushes are expanded into a multi-lobed pattern by projecting spin–orbit beams with varying OAM onto the retina. The number and location of lobes depend on the OAM value ($\ell$): when $\ell < 2$, the modulation appears outside the stimulus ring, while for $\ell > 2$, it shifts inward. Through psychophysical measurements across retinal eccentricities from $0.5^\circ$ to $4^\circ$, we found that perceptual thresholds decrease (improve) with increasing eccentricity, consistent with a scattering-based mechanism involving isotropically distributed retinal structures. Our results reveal a previously untapped class of topologically structured entoptic phenomena and suggest new tools for noninvasive retinal diagnostics, precision assessment of peripheral polarization sensitivity, and controlled studies of light–matter coupling in human vision.

\begin{figure*}
    \centering
    \includegraphics[width=\linewidth]{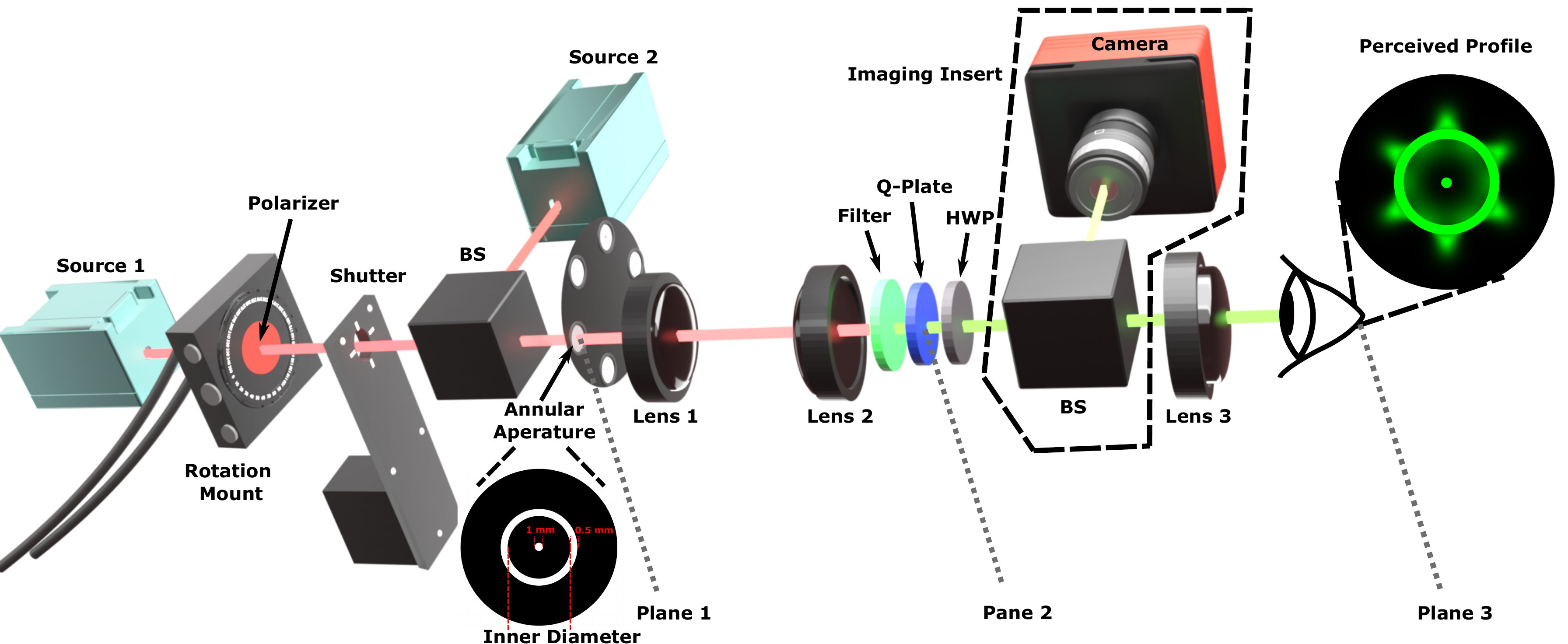}
   \caption{Experimental setup for projecting structured stimuli onto the retina. Two broadband white light sources were combined at a beamsplitter (BS); one arm included a rotating polarizer and shutter, while the other served as a DC offset. This allowed modulation of polarization contrast at fixed intensity. The combined beam passed through an annular aperture and was imaged onto a Q-plate to generate a spin–orbit beam with OAM $\ell = +4$, which was flipped to $\ell = -4$ by a half-wave plate (HWP). The Q-plate output was imaged onto the retina using a f=$125$ mm lens placed in front of the eye. An imaging insert before the final lens enabled alignment with fundus photos to calibrate aperture radius to retinal eccentricity. A 530~nm filter set the stimulus wavelength.}

    \label{fig:fig2}
\end{figure*}

\section{Polarization-Sensitive Scattering in the Retina}

Entoptic phenomena such as Haidinger’s and Boehm’s brushes arise from interactions between polarized light and structures within the human retina. Haidinger’s brushes, visible in central vision, are attributed to the dichroic absorption of short-wavelength light by macular pigments bound perpendicularly to radially oriented Henle fibers. In contrast, Boehm’s brushes appear as faint patterns outside a small, centrally fixated, polarized point source (see Fig.~\ref{fig:fig1}a) and are thought to result from polarization-sensitive scattering within the layered microstructure of the retina.

While Mie scattering within the retina is the underlying physical mechanism, a rigorous quantitative model linking scattering anisotropy and polarization to the perceived entoptic pattern has yet to be established. For the present analysis, we therefore adopt a phenomenological description that captures the observed spatial form of Boehm’s brushes:

\begin{align}
I(r,\phi) \propto \exp\left[-\frac{r^2}{2\sigma_1^2}\right] \cdot \exp\left[-\frac{1}{2\sigma_2^2} \cos^2\left(\phi-\theta \right)\right],
\end{align}

\noindent where $\sigma_1$ characterizes the radial extent, $\sigma_2$ the angular sharpness of the lobes, $(r,\phi)$ are the cylindrical coordinates, and $\theta$ is the polarization orientation.

In our experiments, spin–orbit structured light was used to drive this scattering-mediated perceptual channel. These structured light beams combine circular polarization with OAM, giving rise to azimuthally varying polarization topologies, see Fig.~\ref{fig:fig1}b. The transverse wavefunction of a spin–orbit beam propagating along the $z$-axis can be expressed as:

\begin{align}
	\ket{\Psi} = f(r)\left[e^{i\ell\phi}\ket{R} + \ket{L}\right],
	\label{Eqn:PsiInGeneral}
\end{align}

\noindent where $f(r)$ describes the profile of the stimulus aperture, which in our experiments was an annular profile, see Fig.~\ref{fig:fig2}; $\ket{R} = \begin{pmatrix} 1 \\ 0 \end{pmatrix}$ and $\ket{L} = \begin{pmatrix} 0 \\ 1 \end{pmatrix}$ denote right- and left-circular polarization states, respectively, and $\ell$ is the OAM number. 

The perceived pattern corresponds to the incoherent superposition of locally polarized scattering responses (individual Boehm’s brushes) oriented by the spatially varying polarization of Eq.~\eqref{Eqn:PsiInGeneral}. This yields a global entoptic structure with azimuthal modulation, producing a multi-lobed pattern. The resulting percepts, shown in Fig.~\ref{fig:fig1}b, feature a number of bright lobes determined by:

\begin{equation}
N = |\ell - 2|,
\end{equation}

\noindent reflecting the topological structure of the polarization field. Interestingly, the same lobe-counting relation is observed in previous reports of spin–orbit–modulated Haidinger’s brushes~\cite{sarenac2020direct}.

A notable and unexpected feature of this phenomenon is that the entoptic lobes can appear either inside or outside the visible stimulus region. For $\ell < 2$, the constructive intensity amplification is away from the stimuli, resulting in outer lobes. Conversely, for $\ell > 2$, the constructive intensity leads to lobes perceived within the inner region of the annular stimulus. This behavior is illustrated in Fig.~\ref{fig:fig1}b, which shows how the apparent lobe location depends on the OAM value. The perceived lobe geometry is invariant under changes in eye position: shifting fixation does not alter the number or orientation of lobes. However, the perceived contrast varies across the retina, reflecting local differences in the density and scattering properties of subcellular structures.

\begin{figure*}
    \centering
    \includegraphics[width=\linewidth]{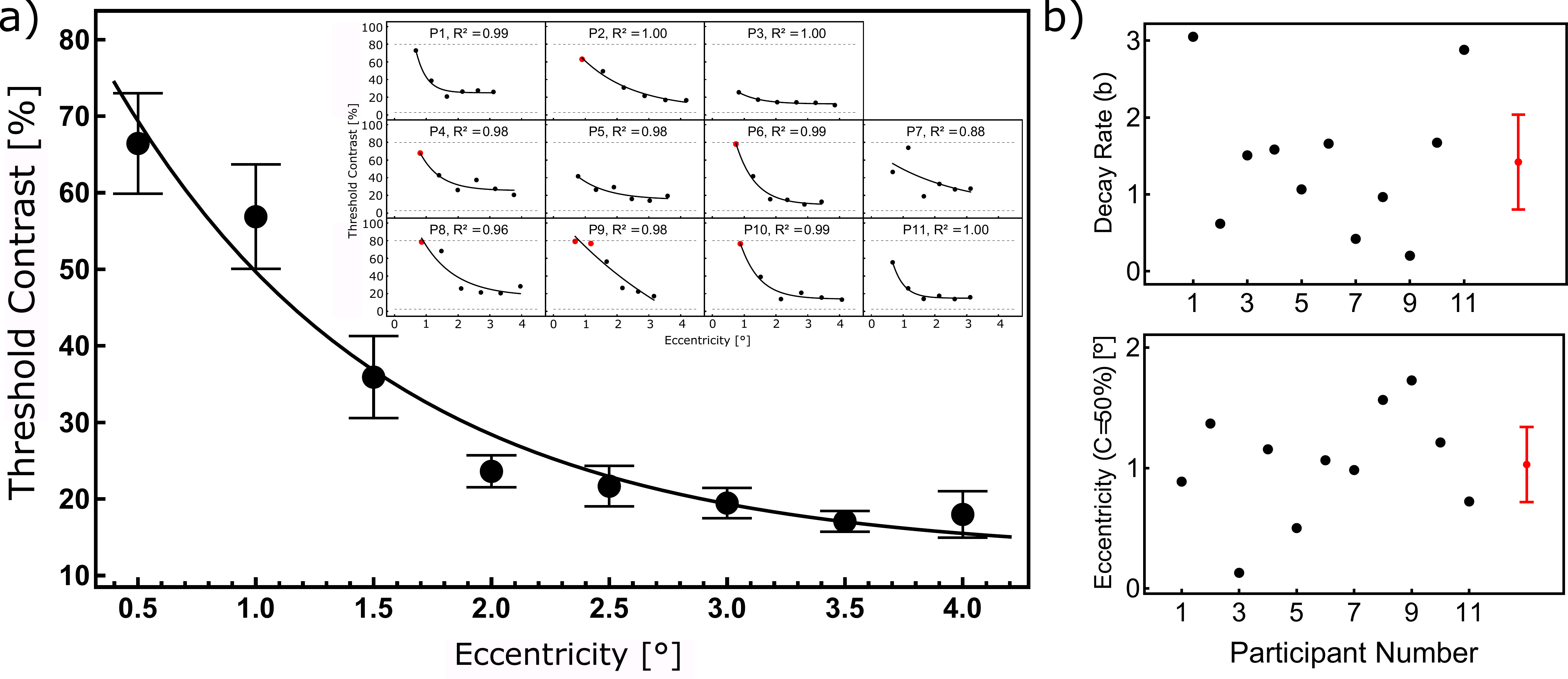}
     \caption{(a) Group-averaged contrast thresholds for entoptic pattern detection across retinal eccentricity. Black points indicate the mean threshold across participants (11 total), where each point represents the average of data within eccentricity bins centered at 0.5$^\circ$, 1.0$^\circ$, 1.5$^\circ$, 2.0$^\circ$, 2.5$^\circ$, 3.0$^\circ$, 3.5$^\circ$, and 4$^\circ$ with a bin width of ±0.25$^\circ$. The solid curve shows the best-fit exponential function, $T(r) = 0.13 + 0.87\, e^{-0.85\, r}$, which captures the group-averaged decay in contrast threshold with increasing eccentricity. The inset shows the individual participant data and fits, illustrating the consistency of the exponential decay across subjects. The red points denote instances where the staircase reached the ceiling (maximum contrast) in two or more of the last six reversals. Note that these represent underestimates of the true threshold. The stimulus contrast range in the setup spanned from 80\% to 2.5\%. (b) Summary of fitted model parameters across participants. The upper panel shows the distribution of decay constants ($b$) extracted from the individual fits, with a group mean of $b = 1.42~\mathrm{deg}^{-1}$ and a 95\% confidence interval of [0.80, 2.04]$~\mathrm{deg}^{-1}$. The lower panel shows the eccentricity at which the fitted models reached a contrast threshold of 50\%, with a mean value of $r_{50} = 1.03^\circ$ and a 95\% confidence interval of [0.72, 1.34]$^\circ$. These results indicate that the entoptic polarization pattern becomes perceptually robust at approximately 1$^\circ$ retinal eccentricity. Across all participants, thresholds were highest near the fovea and decreased (improved) with increasing retinal eccentricity, consistent with a polarization-sensitive scattering mechanism in the retina.}
    \label{fig:fig3}
\end{figure*}

\section{Methods}

\subsection{Experimental Setup}

The experimental system, depicted in Fig.~\ref{fig:fig2}, was designed to deliver spin–orbit coupled light beams with tunable spatial structure to the human retina via an annular aperture. All optical components were aligned to maintain polarization fidelity and spatial mode purity at the retina of the eye. The optical stimulus was generated using two intensity-variable white light sources directed into a non-polarizing 50:50 beamsplitter. One input arm contained a rotating linear polarizer and delivered a polarized beam, while the other arm delivered an unpolarized beam. By balancing the relative intensities of the polarized and unpolarized components, the contrast of the structured stimulus was precisely controlled while keeping the total light intensity constant. A mechanical shutter after the polarizer was used to define the stimulus presentation window, ensuring a precise 250~ms exposure time independent of the motor’s acceleration or deceleration phases.

The importance of rotation speed was established in preliminary trials, which showed that faster or slower spin–orbit beam rotation effectively stretched or compressed the threshold–eccentricity curve. Based on pilot data, polarizer rotation speed 540$^\circ$/s was chosen to capture the contrast threshold decay behavior in typical participants. Note that every 180$^\circ$/s rotation of the polarizer corresponds to one full period of rotation for all entoptic patterns depicted in Fig.~\ref{fig:fig1}b. For the six-lobed structure shown in Fig.~\ref{fig:fig2}, this represents a physical rotation of 60$^\circ$/s; therefore, a polarizer rotation speed of 540$^\circ$/s corresponds to an effective 180$^\circ$/s physical rotation of the perceived pattern.

The output of the beamsplitter was directed through an interchangeable annular aperture. Six annular apertures were used, each with a 0.5 mm-wide opening and inner diameters of 3.9, 7.1, 10.3, 13.5, 16.7, and 19.9 mm. For data analysis, the corresponding retinal eccentricities were defined by the midpoint radius of each annulus, i.e., the average of its inner and outer edges. The geometry of the annular aperture is illustrated in Fig.~\ref{fig:fig2}. A separate 1 mm central circular opening served as the fixation point and was covered with transparent tape to depolarize the light in that region. The aperture plane (Plane 1) was imaged onto a Q-plate using a 4f imaging system consisting of two lenses each with a focal distance of 100 mm. The Q-plate is a liquid crystal optical element that can form a light beam with OAM given a polarized input~\cite{marrucci2006optical}. In our case the Q-plate was optimized to generate a spin-orbit beam with OAM value of $\ell=4$ for 532~nm wavelength light. A half-wave plate placed after the Q-plate changed the circular polarization state to which the OAM was coupled. This adjustment changed the OAM component as written in Eq.~1 to \(\ell = -4\), resulting in an entoptic pattern with six outer azimuthal lobes instead of two inner lobes.

The state after the Q-plate (Plane 2) was then imaged onto the retina of the participant using a f=$125$ mm lens positioned directly in front of the eye. The three lenses in the setup were selected to ensure that the annular apertures spanned an eccentricity range of approximately $0.5^\circ$ to $4^\circ$.

To accurately calibrate the visual eccentricity of each annular aperture, an imaging insert was incorporated into the setup. This insert consisted of a beamsplitter, an imaging lens, and a camera, enabling retinal imaging under illumination by the setup. Retinal images were acquired through the camera system and later compared to participant-specific fundus photographs, enabling extraction of the annular aperture radius–to–retinal eccentricity conversion for each individual. 

A narrowband optical filter centered at 530 nm with a full width at half maximum (FWHM) of 10 nm was placed before the Q-plate to ensure a well-defined wavelength, selected for its high perceptual sensitivity to polarization-dependent scattering~\cite{weale1976spectral} and minimal macular pigment absorption which would enhance visibility of the Haidinger's brushes phenomenon~\cite{bone1980role}.

\subsection{Participants}
All participants were recruited at the University of Waterloo. Participants provided written informed consent to take part in the study. The study protocol was approved by the institutional ethics committee at the University of Waterloo, in accordance with the Declaration of Helsinki. All participants were naive to the experimental hypothesis and remunerated for their time. All tests were performed on the right eye only. Clinical screening included: habitual visual acuity using an Early Treatment of Diabetic Retinopathy Study (ETDRS) chart calibrated for 4 m, objective refractive error determined by an autorefractor (Topcon KR-1, Tokyo, Japan), structural macular integrity by optical coherence tomography (Topcon 3D OCT-1, Tokyo, Japan) and macular luminance sensitivity by microperimetry (4-2 (fast) threshold strategy for central 8 degrees; Nidek MP-3, Aichi, Japan). Fundus imaging was performed undilated with the Nidek MP-3. Participants were excluded if they had visual acuity poorer than 0.2 logMAR, any abnormality visible on macular OCT or an absolute loss of sensitivity on any tested point on microperimetry.

\subsection{Psychophysical Procedure}

Participants completed two structured-light testing sessions following routine optometric screening to verify normal or corrected-to-normal visual acuity and binocular alignment. Session 1 included retinal imaging and a familiarization task to assess perceptual visibility of the structured stimulus. Retinal images were acquired using the same optical system equipped with an imaging insert, capturing views through a calibration aperture. These images were compared with fundus photographs to map aperture geometry to visual angle, allowing subject-specific linkage between annular aperture size and retinal eccentricity.

The familiarization task presented high-contrast clockwise and counterclockwise rotating stimuli. Participants then completed ten randomized trials at fixed contrast to verify comprehension of the task and alignment within the apparatus. Only those achieving at least 8 out of 10 correct responses were advanced to the main contrast-thresholding task.

Session 2 involved quantifying perceptual sensitivity using a two-alternative forced-choice (2AFC) contrast detection task with a 2-up/1-down staircase procedure, converging at 70.7 \% accuracy. A detailed description of the protocol for session 2 is given in the Appendix. Each annular aperture condition was tested using two interleaved staircases: Staircase A began at a contrast of 80 \%, and Staircase B at 30 \%. Each staircase consisted of 14 reversals (maximum 90 trials), with step sizes decreasing across four stages: 0–2 reversals, 0.1; 3–5 reversals, 0.05; 6–8 reversals, 0.025; and 9–14 reversals, 0.0125. Individual contrast thresholds were computed as the arithmetic mean of the final six reversal points.

The structured light stimulus was presented for 250~ms, after which the polarized light path was closed. Participants then reported the perceived direction of azimuthal rotation (clockwise or counterclockwise). Contrast decreased following two consecutive correct responses and increased after a single incorrect response. The response to each trial triggered the onset of the next one. Between aperture conditions, participants were offered a short break lasting a few minutes.

The full protocol consisted of six randomized aperture conditions, corresponding approximately to retinal eccentricities between 0.5$^\circ$ and 4$^\circ$, with presentation order counterbalanced across participants. These eccentricities were selected because Boehm’s brushes are typically absent at the fovea and most prominent around 4$^\circ$, making this range optimal for probing its perceptual onset and peripheral enhancement.

Sixteen participants were initially enrolled. One participant withdrew from the study before scheduling Session 2, and another was found ineligible during the familiarization task of Session 2. Two participants were unable to complete the structured-light retinal imaging due to technical difficulties and did not proceed to Session 2. One additional participant completed both sessions but expressed discomfort and exhibited no measurable sensitivity, with all thresholds approaching the maximum contrast limit. This participant was excluded from further analysis. The remaining eleven participants successfully completed the full protocol and were included in the results shown in Fig.~\ref{fig:fig3}.

\section{Results and Discussion}

Contrast detection thresholds consistently decreased (improved) with increasing retinal eccentricity, see Fig.~\ref{fig:fig3}. Visual inspection of the data in Fig.~\ref{fig:fig3}a revealed a steep initial decline in contrast threshold near the fovea followed by a gradual asymptote at larger eccentricities. This pattern is characteristic of an exponential decay and therefore we can model each participant’s data using the function

\begin{equation}
    T(r) = ae^{-br} + c
    \label{eqn:fit}
\end{equation}

\noindent where \(T(r)\) is the contrast threshold at eccentricity \(r\), and \(a, b, c\) are fit parameters. Each participant’s data were fit individually using this model, yielding high coefficients of determination (R$^2$) across all subjects (inset of Fig.~\ref{fig:fig3}a), confirming that the exponential function captures the observed relationship between threshold and eccentricity. 

Figure~\ref{fig:fig3}a shows the group-averaged thresholds, obtained by binning data within eccentricity intervals centered at 0.5$^\circ$, 1.0$^\circ$, 1.5$^\circ$, 2.0$^\circ$, 2.5$^\circ$, 3.0$^\circ$, 3.5$^\circ$, and 4$^\circ$ (bin width ±0.25$^\circ$). The solid curve represents the best-fit exponential to these binned data, $T(r) = 0.13 + 0.87\, e^{-0.85\, r}$, which reproduces the overall decline in threshold with increasing eccentricity. The inset illustrates individual fits for all participants, demonstrating consistent exponential behavior across subjects. 

The distribution of fitted decay constants (\(b\)) across participants is shown in the top panel of Fig.~\ref{fig:fig3}b, with a mean value of (b = 1.42$~\mathrm{deg}^{-1}$ and a 95\% confidence interval of [0.80, 2.04]$~\mathrm{deg}^{-1}$. The lower panel shows the eccentricities at which the fitted curves reached a 50\% contrast threshold, with a mean value of \(r_{50} = 1.03^\circ\) and a 95\% confidence interval of [0.72, 1.34]$^\circ$. These results indicate that the entoptic pattern becomes perceptually robust at approximately 1$^\circ$ retinal eccentricity. 

While individual thresholds varied, all participants exhibited the same qualitative decay behavior, supporting a common underlying mechanism. Variation in absolute sensitivity may reflect anatomical differences in the density, distribution, or optical properties of retinal scatterers. The consistent decline in threshold with eccentricity supports the conclusion that this structured entoptic phenomenon arises from peripheral, polarization-dependent scattering—distinct from macula-confined absorption-based effects such as Haidinger’s brushes.

\section{Conclusion}

We have shown that structured light carrying spin–orbit coupling elicits a previously unreported entoptic phenomenon in which the classical two-lobed Boehm’s brushes expand into a multi-lobed pattern. The appearance and geometry of these lobes arise from polarization-sensitive scattering in the retina and vary systematically with the topological charge and polarization structure of the incident beam. Psychophysical measurements revealed an exponential decline in detection threshold with increasing retinal eccentricity, with the entoptic response becoming robustly visible near 1$^\circ$ eccentricity. Together, these results demonstrate that topologically structured polarization fields can directly modulate scattering-based entoptic perception, providing a new experimental route for screening retinal integrity, probing peripheral visual sensitivity, and investigating the interaction between structured light and human vision.

Possible future directions include developing quantitative models of the entoptic response using contrast sensitivity functions adapted for polarization-dependent scattering, analogous to those formulated for absorption-based entoptic phenomena~\cite{pushin2025characterizing}. Such models could be compared with OCT-derived retinal thickness profiles to evaluate potential correlations between perceptual thresholds and retinal microstructure. In addition, systematic investigations could explore how retinal structure and visual sensitivity influence the perception of polarization-structured light, for example by varying stimulus rotation speed, illumination wavelength, and retinal eccentricity range.

\section*{Acknowledgments}

This work was supported by the Natural Sciences and Engineering Research Council of Canada (NSERC) grants [RGPIN$-2018-04989$], [RPIN$-05394$], [RGPAS$-477166$],  the Government of Canada’s New Frontiers in Research Fund (NFRF) [NFRFE$-2019-00446$], and the Canada  First  Research  Excellence  Fund  (CFREF). This study was supported by the InnoHK initiative of the Innovation and Technology Commission of the Hong Kong Special Administrative Region Government.

\bibliography{OAM}

\begin{thebibliography}{34}%
\makeatletter
\providecommand \@ifxundefined [1]{%
 \@ifx{#1\undefined}
}%
\providecommand \@ifnum [1]{%
 \ifnum #1\expandafter \@firstoftwo
 \else \expandafter \@secondoftwo
 \fi
}%
\providecommand \@ifx [1]{%
 \ifx #1\expandafter \@firstoftwo
 \else \expandafter \@secondoftwo
 \fi
}%
\providecommand \natexlab [1]{#1}%
\providecommand \enquote  [1]{``#1''}%
\providecommand \bibnamefont  [1]{#1}%
\providecommand \bibfnamefont [1]{#1}%
\providecommand \citenamefont [1]{#1}%
\providecommand \href@noop [0]{\@secondoftwo}%
\providecommand \href [0]{\begingroup \@sanitize@url \@href}%
\providecommand \@href[1]{\@@startlink{#1}\@@href}%
\providecommand \@@href[1]{\endgroup#1\@@endlink}%
\providecommand \@sanitize@url [0]{\catcode `\\12\catcode `\$12\catcode `\&12\catcode `\#12\catcode `\^12\catcode `\_12\catcode `\%12\relax}%
\providecommand \@@startlink[1]{}%
\providecommand \@@endlink[0]{}%
\providecommand \url  [0]{\begingroup\@sanitize@url \@url }%
\providecommand \@url [1]{\endgroup\@href {#1}{\urlprefix }}%
\providecommand \urlprefix  [0]{URL }%
\providecommand \Eprint [0]{\href }%
\providecommand \doibase [0]{http://dx.doi.org/}%
\providecommand \selectlanguage [0]{\@gobble}%
\providecommand \bibinfo  [0]{\@secondoftwo}%
\providecommand \bibfield  [0]{\@secondoftwo}%
\providecommand \translation [1]{[#1]}%
\providecommand \BibitemOpen [0]{}%
\providecommand \bibitemStop [0]{}%
\providecommand \bibitemNoStop [0]{.\EOS\space}%
\providecommand \EOS [0]{\spacefactor3000\relax}%
\providecommand \BibitemShut  [1]{\csname bibitem#1\endcsname}%
\let\auto@bib@innerbib\@empty
\bibitem [{\citenamefont {Rubinsztein-Dunlop}\ \emph {et~al.}(2016)\citenamefont {Rubinsztein-Dunlop}, \citenamefont {Forbes}, \citenamefont {Berry}, \citenamefont {Dennis}, \citenamefont {Andrews}, \citenamefont {Mansuripur}, \citenamefont {Denz}, \citenamefont {Alpmann}, \citenamefont {Banzer}, \citenamefont {Bauer} \emph {et~al.}}]{rubinsztein2016roadmap}%
  \BibitemOpen
  \bibfield  {author} {\bibinfo {author} {\bibfnamefont {Halina}\ \bibnamefont {Rubinsztein-Dunlop}}, \bibinfo {author} {\bibfnamefont {Andrew}\ \bibnamefont {Forbes}}, \bibinfo {author} {\bibfnamefont {MV}~\bibnamefont {Berry}}, \bibinfo {author} {\bibfnamefont {MR}~\bibnamefont {Dennis}}, \bibinfo {author} {\bibfnamefont {David~L}\ \bibnamefont {Andrews}}, \bibinfo {author} {\bibfnamefont {Masud}\ \bibnamefont {Mansuripur}}, \bibinfo {author} {\bibfnamefont {Cornelia}\ \bibnamefont {Denz}}, \bibinfo {author} {\bibfnamefont {Christina}\ \bibnamefont {Alpmann}}, \bibinfo {author} {\bibfnamefont {Peter}\ \bibnamefont {Banzer}}, \bibinfo {author} {\bibfnamefont {Thomas}\ \bibnamefont {Bauer}},  \emph {et~al.},\ }\bibfield  {title} {\enquote {\bibinfo {title} {Roadmap on structured light},}\ }\href@noop {} {\bibfield  {journal} {\bibinfo  {journal} {Journal of Optics}\ }\textbf {\bibinfo {volume} {19}},\ \bibinfo {pages} {013001} (\bibinfo {year} {2016})}\BibitemShut {NoStop}%
\bibitem [{\citenamefont {Bliokh}\ \emph {et~al.}(2023)\citenamefont {Bliokh}, \citenamefont {Karimi}, \citenamefont {Padgett}, \citenamefont {Alonso}, \citenamefont {Dennis}, \citenamefont {Dudley}, \citenamefont {Forbes}, \citenamefont {Zahedpour}, \citenamefont {Hancock}, \citenamefont {Milchberg} \emph {et~al.}}]{bliokh2023roadmap}%
  \BibitemOpen
  \bibfield  {author} {\bibinfo {author} {\bibfnamefont {Konstantin~Y}\ \bibnamefont {Bliokh}}, \bibinfo {author} {\bibfnamefont {Ebrahim}\ \bibnamefont {Karimi}}, \bibinfo {author} {\bibfnamefont {Miles~J}\ \bibnamefont {Padgett}}, \bibinfo {author} {\bibfnamefont {Miguel~A}\ \bibnamefont {Alonso}}, \bibinfo {author} {\bibfnamefont {Mark~R}\ \bibnamefont {Dennis}}, \bibinfo {author} {\bibfnamefont {Angela}\ \bibnamefont {Dudley}}, \bibinfo {author} {\bibfnamefont {Andrew}\ \bibnamefont {Forbes}}, \bibinfo {author} {\bibfnamefont {Sina}\ \bibnamefont {Zahedpour}}, \bibinfo {author} {\bibfnamefont {Scott~W}\ \bibnamefont {Hancock}}, \bibinfo {author} {\bibfnamefont {Howard~M}\ \bibnamefont {Milchberg}},  \emph {et~al.},\ }\bibfield  {title} {\enquote {\bibinfo {title} {Roadmap on structured waves},}\ }\href@noop {} {\bibfield  {journal} {\bibinfo  {journal} {Journal of Optics}\ }\textbf {\bibinfo {volume} {25}},\ \bibinfo {pages} {103001} (\bibinfo {year} {2023})}\BibitemShut {NoStop}%
\bibitem [{\citenamefont {Chen}\ \emph {et~al.}(2021)\citenamefont {Chen}, \citenamefont {Wan},\ and\ \citenamefont {Zhan}}]{chen2021engineering}%
  \BibitemOpen
  \bibfield  {author} {\bibinfo {author} {\bibfnamefont {Jian}\ \bibnamefont {Chen}}, \bibinfo {author} {\bibfnamefont {Chenhao}\ \bibnamefont {Wan}}, \ and\ \bibinfo {author} {\bibfnamefont {Qiwen}\ \bibnamefont {Zhan}},\ }\bibfield  {title} {\enquote {\bibinfo {title} {Engineering photonic angular momentum with structured light: a review},}\ }\href@noop {} {\bibfield  {journal} {\bibinfo  {journal} {Advanced Photonics}\ }\textbf {\bibinfo {volume} {3}},\ \bibinfo {pages} {064001--064001} (\bibinfo {year} {2021})}\BibitemShut {NoStop}%
\bibitem [{\citenamefont {Ni}\ \emph {et~al.}(2021)\citenamefont {Ni}, \citenamefont {Huang}, \citenamefont {Zhou}, \citenamefont {Gu}, \citenamefont {Song}, \citenamefont {Kivshar},\ and\ \citenamefont {Qiu}}]{ni2021multidimensional}%
  \BibitemOpen
  \bibfield  {author} {\bibinfo {author} {\bibfnamefont {Jincheng}\ \bibnamefont {Ni}}, \bibinfo {author} {\bibfnamefont {Can}\ \bibnamefont {Huang}}, \bibinfo {author} {\bibfnamefont {Lei-Ming}\ \bibnamefont {Zhou}}, \bibinfo {author} {\bibfnamefont {Min}\ \bibnamefont {Gu}}, \bibinfo {author} {\bibfnamefont {Qinghai}\ \bibnamefont {Song}}, \bibinfo {author} {\bibfnamefont {Yuri}\ \bibnamefont {Kivshar}}, \ and\ \bibinfo {author} {\bibfnamefont {Cheng-Wei}\ \bibnamefont {Qiu}},\ }\bibfield  {title} {\enquote {\bibinfo {title} {Multidimensional phase singularities in nanophotonics},}\ }\href@noop {} {\bibfield  {journal} {\bibinfo  {journal} {Science}\ }\textbf {\bibinfo {volume} {374}},\ \bibinfo {pages} {eabj0039} (\bibinfo {year} {2021})}\BibitemShut {NoStop}%
\bibitem [{\citenamefont {Allen}\ \emph {et~al.}(1992)\citenamefont {Allen}, \citenamefont {Beijersbergen}, \citenamefont {Spreeuw},\ and\ \citenamefont {Woerdman}}]{allen1992orbital}%
  \BibitemOpen
  \bibfield  {author} {\bibinfo {author} {\bibfnamefont {Les}\ \bibnamefont {Allen}}, \bibinfo {author} {\bibfnamefont {Marco~W}\ \bibnamefont {Beijersbergen}}, \bibinfo {author} {\bibfnamefont {RJC}\ \bibnamefont {Spreeuw}}, \ and\ \bibinfo {author} {\bibfnamefont {JP}~\bibnamefont {Woerdman}},\ }\bibfield  {title} {\enquote {\bibinfo {title} {Orbital angular momentum of light and the transformation of laguerre-gaussian laser modes},}\ }\href@noop {} {\bibfield  {journal} {\bibinfo  {journal} {Physical review A}\ }\textbf {\bibinfo {volume} {45}},\ \bibinfo {pages} {8185} (\bibinfo {year} {1992})}\BibitemShut {NoStop}%
\bibitem [{\citenamefont {Shen}\ \emph {et~al.}(2019)\citenamefont {Shen}, \citenamefont {Wang}, \citenamefont {Xie}, \citenamefont {Min}, \citenamefont {Fu}, \citenamefont {Liu}, \citenamefont {Gong},\ and\ \citenamefont {Yuan}}]{shen2019optical}%
  \BibitemOpen
  \bibfield  {author} {\bibinfo {author} {\bibfnamefont {Yijie}\ \bibnamefont {Shen}}, \bibinfo {author} {\bibfnamefont {Xuejiao}\ \bibnamefont {Wang}}, \bibinfo {author} {\bibfnamefont {Zhenwei}\ \bibnamefont {Xie}}, \bibinfo {author} {\bibfnamefont {Changjun}\ \bibnamefont {Min}}, \bibinfo {author} {\bibfnamefont {Xing}\ \bibnamefont {Fu}}, \bibinfo {author} {\bibfnamefont {Qiang}\ \bibnamefont {Liu}}, \bibinfo {author} {\bibfnamefont {Mali}\ \bibnamefont {Gong}}, \ and\ \bibinfo {author} {\bibfnamefont {Xiaocong}\ \bibnamefont {Yuan}},\ }\bibfield  {title} {\enquote {\bibinfo {title} {Optical vortices 30 years on: Oam manipulation from topological charge to multiple singularities},}\ }\href@noop {} {\bibfield  {journal} {\bibinfo  {journal} {Light: Science \& Applications}\ }\textbf {\bibinfo {volume} {8}},\ \bibinfo {pages} {90} (\bibinfo {year} {2019})}\BibitemShut {NoStop}%
\bibitem [{\citenamefont {Mair}\ \emph {et~al.}(2001)\citenamefont {Mair}, \citenamefont {Vaziri}, \citenamefont {Weihs},\ and\ \citenamefont {Zeilinger}}]{mair2001entanglement}%
  \BibitemOpen
  \bibfield  {author} {\bibinfo {author} {\bibfnamefont {Alois}\ \bibnamefont {Mair}}, \bibinfo {author} {\bibfnamefont {Alipasha}\ \bibnamefont {Vaziri}}, \bibinfo {author} {\bibfnamefont {Gregor}\ \bibnamefont {Weihs}}, \ and\ \bibinfo {author} {\bibfnamefont {Anton}\ \bibnamefont {Zeilinger}},\ }\bibfield  {title} {\enquote {\bibinfo {title} {Entanglement of the orbital angular momentum states of photons},}\ }\href@noop {} {\bibfield  {journal} {\bibinfo  {journal} {Nature}\ }\textbf {\bibinfo {volume} {412}},\ \bibinfo {pages} {313--316} (\bibinfo {year} {2001})}\BibitemShut {NoStop}%
\bibitem [{\citenamefont {{Andersen}}\ \emph {et~al.}(2006)\citenamefont {{Andersen}}, \citenamefont {{Ryu}}, \citenamefont {{Clad{\'e}}}, \citenamefont {{Natarajan}}, \citenamefont {{Vaziri}}, \citenamefont {{Helmerson}},\ and\ \citenamefont {{Phillips}}}]{Andersen2006}%
  \BibitemOpen
  \bibfield  {author} {\bibinfo {author} {\bibfnamefont {M.~F.}\ \bibnamefont {{Andersen}}}, \bibinfo {author} {\bibfnamefont {C.}~\bibnamefont {{Ryu}}}, \bibinfo {author} {\bibfnamefont {P.}~\bibnamefont {{Clad{\'e}}}}, \bibinfo {author} {\bibfnamefont {V.}~\bibnamefont {{Natarajan}}}, \bibinfo {author} {\bibfnamefont {A.}~\bibnamefont {{Vaziri}}}, \bibinfo {author} {\bibfnamefont {K.}~\bibnamefont {{Helmerson}}}, \ and\ \bibinfo {author} {\bibfnamefont {W.~D.}\ \bibnamefont {{Phillips}}},\ }\bibfield  {title} {\enquote {\bibinfo {title} {{Quantized Rotation of Atoms from Photons with Orbital Angular Momentum}},}\ }\href {\doibase 10.1103/PhysRevLett.97.170406} {\bibfield  {journal} {\bibinfo  {journal} {Physical Review Letters}\ }\textbf {\bibinfo {volume} {97}},\ \bibinfo {eid} {170406} (\bibinfo {year} {2006})}\BibitemShut {NoStop}%
\bibitem [{\citenamefont {Cameron}\ \emph {et~al.}(2021)\citenamefont {Cameron}, \citenamefont {Cheng}, \citenamefont {Schwarz}, \citenamefont {Kapahi}, \citenamefont {Sarenac}, \citenamefont {Grabowecky}, \citenamefont {Cory}, \citenamefont {Jennewein}, \citenamefont {Pushin},\ and\ \citenamefont {Resch}}]{cameron2021remote}%
  \BibitemOpen
  \bibfield  {author} {\bibinfo {author} {\bibfnamefont {Andrew~R}\ \bibnamefont {Cameron}}, \bibinfo {author} {\bibfnamefont {Sandra~WL}\ \bibnamefont {Cheng}}, \bibinfo {author} {\bibfnamefont {Sacha}\ \bibnamefont {Schwarz}}, \bibinfo {author} {\bibfnamefont {Connor}\ \bibnamefont {Kapahi}}, \bibinfo {author} {\bibfnamefont {Dusan}\ \bibnamefont {Sarenac}}, \bibinfo {author} {\bibfnamefont {Michael}\ \bibnamefont {Grabowecky}}, \bibinfo {author} {\bibfnamefont {David~G}\ \bibnamefont {Cory}}, \bibinfo {author} {\bibfnamefont {Thomas}\ \bibnamefont {Jennewein}}, \bibinfo {author} {\bibfnamefont {Dmitry~A}\ \bibnamefont {Pushin}}, \ and\ \bibinfo {author} {\bibfnamefont {Kevin~J}\ \bibnamefont {Resch}},\ }\bibfield  {title} {\enquote {\bibinfo {title} {Remote state preparation of single-photon orbital-angular-momentum lattices},}\ }\href@noop {} {\bibfield  {journal} {\bibinfo  {journal} {Physical Review A}\ }\textbf {\bibinfo {volume} {104}},\ \bibinfo {pages} {L051701} (\bibinfo {year} {2021})}\BibitemShut
  {NoStop}%
\bibitem [{\citenamefont {Ritsch-Marte}(2017)}]{ritsch2017orbital}%
  \BibitemOpen
  \bibfield  {author} {\bibinfo {author} {\bibfnamefont {Monika}\ \bibnamefont {Ritsch-Marte}},\ }\bibfield  {title} {\enquote {\bibinfo {title} {Orbital angular momentum light in microscopy},}\ }\href@noop {} {\bibfield  {journal} {\bibinfo  {journal} {Philosophical Transactions of the Royal Society A: Mathematical, Physical and Engineering Sciences}\ }\textbf {\bibinfo {volume} {375}},\ \bibinfo {pages} {20150437} (\bibinfo {year} {2017})}\BibitemShut {NoStop}%
\bibitem [{\citenamefont {Wang}\ \emph {et~al.}(2012)\citenamefont {Wang}, \citenamefont {Yang}, \citenamefont {Fazal}, \citenamefont {Ahmed}, \citenamefont {Yan}, \citenamefont {Huang}, \citenamefont {Ren}, \citenamefont {Yue}, \citenamefont {Dolinar}, \citenamefont {Tur} \emph {et~al.}}]{wang2012terabit}%
  \BibitemOpen
  \bibfield  {author} {\bibinfo {author} {\bibfnamefont {Jian}\ \bibnamefont {Wang}}, \bibinfo {author} {\bibfnamefont {Jeng-Yuan}\ \bibnamefont {Yang}}, \bibinfo {author} {\bibfnamefont {Irfan~M}\ \bibnamefont {Fazal}}, \bibinfo {author} {\bibfnamefont {Nisar}\ \bibnamefont {Ahmed}}, \bibinfo {author} {\bibfnamefont {Yan}\ \bibnamefont {Yan}}, \bibinfo {author} {\bibfnamefont {Hao}\ \bibnamefont {Huang}}, \bibinfo {author} {\bibfnamefont {Yongxiong}\ \bibnamefont {Ren}}, \bibinfo {author} {\bibfnamefont {Yang}\ \bibnamefont {Yue}}, \bibinfo {author} {\bibfnamefont {Samuel}\ \bibnamefont {Dolinar}}, \bibinfo {author} {\bibfnamefont {Moshe}\ \bibnamefont {Tur}},  \emph {et~al.},\ }\bibfield  {title} {\enquote {\bibinfo {title} {Terabit free-space data transmission employing orbital angular momentum multiplexing},}\ }\href@noop {} {\bibfield  {journal} {\bibinfo  {journal} {Nature Photonics}\ }\textbf {\bibinfo {volume} {6}},\ \bibinfo {pages} {488--496} (\bibinfo {year} {2012})}\BibitemShut {NoStop}%
\bibitem [{\citenamefont {Maurer}\ \emph {et~al.}(2007)\citenamefont {Maurer}, \citenamefont {Jesacher}, \citenamefont {F{\"u}rhapter}, \citenamefont {Bernet},\ and\ \citenamefont {Ritsch-Marte}}]{maurer2007tailoring}%
  \BibitemOpen
  \bibfield  {author} {\bibinfo {author} {\bibfnamefont {Christian}\ \bibnamefont {Maurer}}, \bibinfo {author} {\bibfnamefont {Alexander}\ \bibnamefont {Jesacher}}, \bibinfo {author} {\bibfnamefont {Severin}\ \bibnamefont {F{\"u}rhapter}}, \bibinfo {author} {\bibfnamefont {Stefan}\ \bibnamefont {Bernet}}, \ and\ \bibinfo {author} {\bibfnamefont {Monika}\ \bibnamefont {Ritsch-Marte}},\ }\bibfield  {title} {\enquote {\bibinfo {title} {Tailoring of arbitrary optical vector beams},}\ }\href@noop {} {\bibfield  {journal} {\bibinfo  {journal} {New Journal of Physics}\ }\textbf {\bibinfo {volume} {9}},\ \bibinfo {pages} {78} (\bibinfo {year} {2007})}\BibitemShut {NoStop}%
\bibitem [{\citenamefont {Sarenac}\ \emph {et~al.}(2018)\citenamefont {Sarenac}, \citenamefont {Cory}, \citenamefont {Nsofini}, \citenamefont {Hincks}, \citenamefont {Miguel}, \citenamefont {Arif}, \citenamefont {Clark}, \citenamefont {Huber},\ and\ \citenamefont {Pushin}}]{sarenac2018generation}%
  \BibitemOpen
  \bibfield  {author} {\bibinfo {author} {\bibfnamefont {D}~\bibnamefont {Sarenac}}, \bibinfo {author} {\bibfnamefont {DG}~\bibnamefont {Cory}}, \bibinfo {author} {\bibfnamefont {J}~\bibnamefont {Nsofini}}, \bibinfo {author} {\bibfnamefont {I}~\bibnamefont {Hincks}}, \bibinfo {author} {\bibfnamefont {P}~\bibnamefont {Miguel}}, \bibinfo {author} {\bibfnamefont {M}~\bibnamefont {Arif}}, \bibinfo {author} {\bibfnamefont {Charles~W}\ \bibnamefont {Clark}}, \bibinfo {author} {\bibfnamefont {MG}~\bibnamefont {Huber}}, \ and\ \bibinfo {author} {\bibfnamefont {DA}~\bibnamefont {Pushin}},\ }\bibfield  {title} {\enquote {\bibinfo {title} {Generation of a lattice of spin-orbit beams via coherent averaging},}\ }\href@noop {} {\bibfield  {journal} {\bibinfo  {journal} {Physical Review Letters}\ }\textbf {\bibinfo {volume} {121}},\ \bibinfo {pages} {183602} (\bibinfo {year} {2018})}\BibitemShut {NoStop}%
\bibitem [{\citenamefont {Marrucci}\ \emph {et~al.}(2006)\citenamefont {Marrucci}, \citenamefont {Manzo},\ and\ \citenamefont {Paparo}}]{marrucci2006optical}%
  \BibitemOpen
  \bibfield  {author} {\bibinfo {author} {\bibfnamefont {Lorenzo}\ \bibnamefont {Marrucci}}, \bibinfo {author} {\bibfnamefont {C}~\bibnamefont {Manzo}}, \ and\ \bibinfo {author} {\bibfnamefont {D}~\bibnamefont {Paparo}},\ }\bibfield  {title} {\enquote {\bibinfo {title} {Optical spin-to-orbital angular momentum conversion in inhomogeneous anisotropic media},}\ }\href@noop {} {\bibfield  {journal} {\bibinfo  {journal} {Physical Review Letters}\ }\textbf {\bibinfo {volume} {96}},\ \bibinfo {pages} {163905} (\bibinfo {year} {2006})}\BibitemShut {NoStop}%
\bibitem [{\citenamefont {Horv{\'a}th}\ and\ \citenamefont {Varju}(2004)}]{horvath2004polarized}%
  \BibitemOpen
  \bibfield  {author} {\bibinfo {author} {\bibfnamefont {G{\'a}bor}\ \bibnamefont {Horv{\'a}th}}\ and\ \bibinfo {author} {\bibfnamefont {Dezs{\"o}}\ \bibnamefont {Varju}},\ }\href@noop {} {\emph {\bibinfo {title} {Polarized light in animal vision: polarization patterns in nature}}}\ (\bibinfo  {publisher} {Springer Science \& Business Media},\ \bibinfo {year} {2004})\BibitemShut {NoStop}%
\bibitem [{\citenamefont {O’Shea}\ \emph {et~al.}(2021)\citenamefont {O’Shea}, \citenamefont {Misson},\ and\ \citenamefont {Temple}}]{o2021seeing}%
  \BibitemOpen
  \bibfield  {author} {\bibinfo {author} {\bibfnamefont {Robert~P}\ \bibnamefont {O’Shea}}, \bibinfo {author} {\bibfnamefont {Gary~P}\ \bibnamefont {Misson}}, \ and\ \bibinfo {author} {\bibfnamefont {Shelby~E}\ \bibnamefont {Temple}},\ }\bibfield  {title} {\enquote {\bibinfo {title} {Seeing polarization of light with the naked eye},}\ }\href@noop {} {\bibfield  {journal} {\bibinfo  {journal} {Current Biology}\ }\textbf {\bibinfo {volume} {31}},\ \bibinfo {pages} {R178--R179} (\bibinfo {year} {2021})}\BibitemShut {NoStop}%
\bibitem [{\citenamefont {Temple}\ and\ \citenamefont {Misson}(2024)}]{temple2024human}%
  \BibitemOpen
  \bibfield  {author} {\bibinfo {author} {\bibfnamefont {Shelby}\ \bibnamefont {Temple}}\ and\ \bibinfo {author} {\bibfnamefont {Gary}\ \bibnamefont {Misson}},\ }\bibfield  {title} {\enquote {\bibinfo {title} {Human polarization sensitivity: An update},}\ }in\ \href@noop {} {\emph {\bibinfo {booktitle} {Polarization Vision and Environmental Polarized Light}}}\ (\bibinfo  {publisher} {Springer},\ \bibinfo {year} {2024})\ pp.\ \bibinfo {pages} {317--345}\BibitemShut {NoStop}%
\bibitem [{\citenamefont {Haidinger}(1844)}]{haidinger1844ueber}%
  \BibitemOpen
  \bibfield  {author} {\bibinfo {author} {\bibfnamefont {Wilhelm}\ \bibnamefont {Haidinger}},\ }\bibfield  {title} {\enquote {\bibinfo {title} {Ueber das directe erkennen des polarisirten lichts und der lage der polarisationsebene},}\ }\href@noop {} {\bibfield  {journal} {\bibinfo  {journal} {Annalen der Physik}\ }\textbf {\bibinfo {volume} {139}},\ \bibinfo {pages} {29--39} (\bibinfo {year} {1844})}\BibitemShut {NoStop}%
\bibitem [{\citenamefont {Misson}\ \emph {et~al.}(2020)\citenamefont {Misson}, \citenamefont {Temple},\ and\ \citenamefont {Anderson}}]{misson2020polarization}%
  \BibitemOpen
  \bibfield  {author} {\bibinfo {author} {\bibfnamefont {Gary~P}\ \bibnamefont {Misson}}, \bibinfo {author} {\bibfnamefont {Shelby~E}\ \bibnamefont {Temple}}, \ and\ \bibinfo {author} {\bibfnamefont {Stephen~J}\ \bibnamefont {Anderson}},\ }\bibfield  {title} {\enquote {\bibinfo {title} {Polarization perception in humans: On the origin of and relationship between maxwell’s spot and haidinger’s brushes},}\ }\href@noop {} {\bibfield  {journal} {\bibinfo  {journal} {Scientific Reports}\ }\textbf {\bibinfo {volume} {10}},\ \bibinfo {pages} {108} (\bibinfo {year} {2020})}\BibitemShut {NoStop}%
\bibitem [{\citenamefont {Mottes}\ \emph {et~al.}(2022)\citenamefont {Mottes}, \citenamefont {Ortolan},\ and\ \citenamefont {Ruffato}}]{mottes2022haidinger}%
  \BibitemOpen
  \bibfield  {author} {\bibinfo {author} {\bibfnamefont {Jacopo}\ \bibnamefont {Mottes}}, \bibinfo {author} {\bibfnamefont {Dominga}\ \bibnamefont {Ortolan}}, \ and\ \bibinfo {author} {\bibfnamefont {Gianluca}\ \bibnamefont {Ruffato}},\ }\bibfield  {title} {\enquote {\bibinfo {title} {Haidinger’s brushes: Psychophysical analysis of an entoptic phenomenon},}\ }\href@noop {} {\bibfield  {journal} {\bibinfo  {journal} {Vision Research}\ }\textbf {\bibinfo {volume} {199}},\ \bibinfo {pages} {108076} (\bibinfo {year} {2022})}\BibitemShut {NoStop}%
\bibitem [{\citenamefont {Misson}\ \emph {et~al.}(2015)\citenamefont {Misson}, \citenamefont {Timmerman},\ and\ \citenamefont {Bryanston-Cross}}]{misson2015human}%
  \BibitemOpen
  \bibfield  {author} {\bibinfo {author} {\bibfnamefont {Gary~P}\ \bibnamefont {Misson}}, \bibinfo {author} {\bibfnamefont {Brenda~H}\ \bibnamefont {Timmerman}}, \ and\ \bibinfo {author} {\bibfnamefont {Peter~J}\ \bibnamefont {Bryanston-Cross}},\ }\bibfield  {title} {\enquote {\bibinfo {title} {Human perception of visual stimuli modulated by direction of linear polarization},}\ }\href@noop {} {\bibfield  {journal} {\bibinfo  {journal} {Vision Research}\ }\textbf {\bibinfo {volume} {115}},\ \bibinfo {pages} {48--57} (\bibinfo {year} {2015})}\BibitemShut {NoStop}%
\bibitem [{\citenamefont {Misson}\ and\ \citenamefont {Anderson}(2017)}]{misson2017spectral}%
  \BibitemOpen
  \bibfield  {author} {\bibinfo {author} {\bibfnamefont {Gary~P}\ \bibnamefont {Misson}}\ and\ \bibinfo {author} {\bibfnamefont {Stephen~J}\ \bibnamefont {Anderson}},\ }\bibfield  {title} {\enquote {\bibinfo {title} {The spectral, spatial and contrast sensitivity of human polarization pattern perception},}\ }\href@noop {} {\bibfield  {journal} {\bibinfo  {journal} {Scientific Reports}\ }\textbf {\bibinfo {volume} {7}},\ \bibinfo {pages} {16571} (\bibinfo {year} {2017})}\BibitemShut {NoStop}%
\bibitem [{\citenamefont {Misson}\ \emph {et~al.}(2019)\citenamefont {Misson}, \citenamefont {Temple},\ and\ \citenamefont {Anderson}}]{misson2019computational}%
  \BibitemOpen
  \bibfield  {author} {\bibinfo {author} {\bibfnamefont {Gary~P}\ \bibnamefont {Misson}}, \bibinfo {author} {\bibfnamefont {Shelby~E}\ \bibnamefont {Temple}}, \ and\ \bibinfo {author} {\bibfnamefont {Stephen~J}\ \bibnamefont {Anderson}},\ }\bibfield  {title} {\enquote {\bibinfo {title} {Computational simulation of human perception of spatially dependent patterns modulated by degree and angle of linear polarization},}\ }\href@noop {} {\bibfield  {journal} {\bibinfo  {journal} {JOSA A}\ }\textbf {\bibinfo {volume} {36}},\ \bibinfo {pages} {B65--B70} (\bibinfo {year} {2019})}\BibitemShut {NoStop}%
\bibitem [{\citenamefont {Wang}\ \emph {et~al.}(2022)\citenamefont {Wang}, \citenamefont {Bryanston-Cross}, \citenamefont {Li},\ and\ \citenamefont {Liu}}]{wang2022mathematical}%
  \BibitemOpen
  \bibfield  {author} {\bibinfo {author} {\bibfnamefont {Qi}~\bibnamefont {Wang}}, \bibinfo {author} {\bibfnamefont {Peter~J}\ \bibnamefont {Bryanston-Cross}}, \bibinfo {author} {\bibfnamefont {Yahong}\ \bibnamefont {Li}}, \ and\ \bibinfo {author} {\bibfnamefont {Zhiying}\ \bibnamefont {Liu}},\ }\bibfield  {title} {\enquote {\bibinfo {title} {Mathematical modeling and experimental verification of aging human eyes polarization sensitivity},}\ }\href@noop {} {\bibfield  {journal} {\bibinfo  {journal} {JOSA A}\ }\textbf {\bibinfo {volume} {39}},\ \bibinfo {pages} {2398--2409} (\bibinfo {year} {2022})}\BibitemShut {NoStop}%
\bibitem [{\citenamefont {Boehm}(1940)}]{boehm1940neues}%
  \BibitemOpen
  \bibfield  {author} {\bibinfo {author} {\bibfnamefont {Gundo}\ \bibnamefont {Boehm}},\ }\bibfield  {title} {\enquote {\bibinfo {title} {{\"U}ber ein neues entoptisches ph{\"a}nomen im polarisierten licht. periphere polarisationsb{\"u}schel},}\ }\href@noop {} {\bibfield  {journal} {\bibinfo  {journal} {Acta Ophthalmologica}\ }\textbf {\bibinfo {volume} {18}},\ \bibinfo {pages} {143--169} (\bibinfo {year} {1940})}\BibitemShut {NoStop}%
\bibitem [{\citenamefont {Vos}\ and\ \citenamefont {Bouman}(1964)}]{vos1964contribution}%
  \BibitemOpen
  \bibfield  {author} {\bibinfo {author} {\bibfnamefont {JJ}~\bibnamefont {Vos}}\ and\ \bibinfo {author} {\bibfnamefont {MA}~\bibnamefont {Bouman}},\ }\bibfield  {title} {\enquote {\bibinfo {title} {Contribution of the retina to entoptic scatter},}\ }\href@noop {} {\bibfield  {journal} {\bibinfo  {journal} {Journal of the Optical Society of America}\ }\textbf {\bibinfo {volume} {54}},\ \bibinfo {pages} {95--100} (\bibinfo {year} {1964})}\BibitemShut {NoStop}%
\bibitem [{\citenamefont {Weale}(1976)}]{weale1976spectral}%
  \BibitemOpen
  \bibfield  {author} {\bibinfo {author} {\bibfnamefont {RA}~\bibnamefont {Weale}},\ }\bibfield  {title} {\enquote {\bibinfo {title} {On the spectral sensitivity of the human retina to light which it has scattered},}\ }\href@noop {} {\bibfield  {journal} {\bibinfo  {journal} {Vision Research}\ }\textbf {\bibinfo {volume} {16}},\ \bibinfo {pages} {1395--1399} (\bibinfo {year} {1976})}\BibitemShut {NoStop}%
\bibitem [{\citenamefont {Sarenac}\ \emph {et~al.}(2020)\citenamefont {Sarenac}, \citenamefont {Kapahi}, \citenamefont {Silva}, \citenamefont {Cory}, \citenamefont {Taminiau}, \citenamefont {Thompson},\ and\ \citenamefont {Pushin}}]{sarenac2020direct}%
  \BibitemOpen
  \bibfield  {author} {\bibinfo {author} {\bibfnamefont {Dusan}\ \bibnamefont {Sarenac}}, \bibinfo {author} {\bibfnamefont {Connor}\ \bibnamefont {Kapahi}}, \bibinfo {author} {\bibfnamefont {Andrew~E}\ \bibnamefont {Silva}}, \bibinfo {author} {\bibfnamefont {David~G}\ \bibnamefont {Cory}}, \bibinfo {author} {\bibfnamefont {Ivar}\ \bibnamefont {Taminiau}}, \bibinfo {author} {\bibfnamefont {Benjamin}\ \bibnamefont {Thompson}}, \ and\ \bibinfo {author} {\bibfnamefont {Dmitry~A}\ \bibnamefont {Pushin}},\ }\bibfield  {title} {\enquote {\bibinfo {title} {Direct discrimination of structured light by humans},}\ }\href@noop {} {\bibfield  {journal} {\bibinfo  {journal} {Proceedings of the National Academy of Sciences}\ }\textbf {\bibinfo {volume} {117}},\ \bibinfo {pages} {14682--14687} (\bibinfo {year} {2020})}\BibitemShut {NoStop}%
\bibitem [{\citenamefont {Sarenac}\ \emph {et~al.}(2022)\citenamefont {Sarenac}, \citenamefont {Silva}, \citenamefont {Kapahi}, \citenamefont {Cory}, \citenamefont {Thompson},\ and\ \citenamefont {Pushin}}]{sarenac2022human}%
  \BibitemOpen
  \bibfield  {author} {\bibinfo {author} {\bibfnamefont {Dusan}\ \bibnamefont {Sarenac}}, \bibinfo {author} {\bibfnamefont {Andrew~E}\ \bibnamefont {Silva}}, \bibinfo {author} {\bibfnamefont {Connor}\ \bibnamefont {Kapahi}}, \bibinfo {author} {\bibfnamefont {DG}~\bibnamefont {Cory}}, \bibinfo {author} {\bibfnamefont {B}~\bibnamefont {Thompson}}, \ and\ \bibinfo {author} {\bibfnamefont {Dmitry~A}\ \bibnamefont {Pushin}},\ }\bibfield  {title} {\enquote {\bibinfo {title} {Human psychophysical discrimination of spatially dependant pancharatnam--berry phases in optical spin-orbit states},}\ }\href@noop {} {\bibfield  {journal} {\bibinfo  {journal} {Scientific Reports}\ }\textbf {\bibinfo {volume} {12}},\ \bibinfo {pages} {3245} (\bibinfo {year} {2022})}\BibitemShut {NoStop}%
\bibitem [{\citenamefont {Kapahi}\ \emph {et~al.}(2024)\citenamefont {Kapahi}, \citenamefont {Silva}, \citenamefont {Cory}, \citenamefont {Kulmaganbetov}, \citenamefont {Mungalsingh}, \citenamefont {Pushin}, \citenamefont {Singh}, \citenamefont {Thompson},\ and\ \citenamefont {Sarenac}}]{kapahi2024measuring}%
  \BibitemOpen
  \bibfield  {author} {\bibinfo {author} {\bibfnamefont {C}~\bibnamefont {Kapahi}}, \bibinfo {author} {\bibfnamefont {AE}~\bibnamefont {Silva}}, \bibinfo {author} {\bibfnamefont {DG}~\bibnamefont {Cory}}, \bibinfo {author} {\bibfnamefont {M}~\bibnamefont {Kulmaganbetov}}, \bibinfo {author} {\bibfnamefont {MA}~\bibnamefont {Mungalsingh}}, \bibinfo {author} {\bibfnamefont {DA}~\bibnamefont {Pushin}}, \bibinfo {author} {\bibfnamefont {T}~\bibnamefont {Singh}}, \bibinfo {author} {\bibfnamefont {B}~\bibnamefont {Thompson}}, \ and\ \bibinfo {author} {\bibfnamefont {D}~\bibnamefont {Sarenac}},\ }\bibfield  {title} {\enquote {\bibinfo {title} {Measuring the visual angle of polarization-related entoptic phenomena using structured light.}}\ }\href@noop {} {\bibfield  {journal} {\bibinfo  {journal} {Biomedical Optics Express}\ }\textbf {\bibinfo {volume} {15}},\ \bibinfo {pages} {1278--1287} (\bibinfo {year} {2024})}\BibitemShut {NoStop}%
\bibitem [{\citenamefont {Pushin}\ \emph {et~al.}(2024)\citenamefont {Pushin}, \citenamefont {Kapahi}, \citenamefont {Silva}, \citenamefont {Cory}, \citenamefont {Kulmaganbetov}, \citenamefont {Mungalsingh}, \citenamefont {Singh}, \citenamefont {Thompson},\ and\ \citenamefont {Sarenac}}]{pushin2024psychophysical}%
  \BibitemOpen
  \bibfield  {author} {\bibinfo {author} {\bibfnamefont {DA}~\bibnamefont {Pushin}}, \bibinfo {author} {\bibfnamefont {C}~\bibnamefont {Kapahi}}, \bibinfo {author} {\bibfnamefont {AE}~\bibnamefont {Silva}}, \bibinfo {author} {\bibfnamefont {DG}~\bibnamefont {Cory}}, \bibinfo {author} {\bibfnamefont {M}~\bibnamefont {Kulmaganbetov}}, \bibinfo {author} {\bibfnamefont {M}~\bibnamefont {Mungalsingh}}, \bibinfo {author} {\bibfnamefont {T}~\bibnamefont {Singh}}, \bibinfo {author} {\bibfnamefont {B}~\bibnamefont {Thompson}}, \ and\ \bibinfo {author} {\bibfnamefont {D}~\bibnamefont {Sarenac}},\ }\bibfield  {title} {\enquote {\bibinfo {title} {Psychophysical discrimination of radially varying polarization-based entoptic phenomena},}\ }\href@noop {} {\bibfield  {journal} {\bibinfo  {journal} {Physical Review Applied}\ }\textbf {\bibinfo {volume} {21}},\ \bibinfo {pages} {L011002} (\bibinfo {year} {2024})}\BibitemShut {NoStop}%
\bibitem [{\citenamefont {Pushin}\ \emph {et~al.}(2023)\citenamefont {Pushin}, \citenamefont {Cory}, \citenamefont {Kapahi}, \citenamefont {Kulmaganbetov}, \citenamefont {Mungalsingh}, \citenamefont {Silva}, \citenamefont {Singh}, \citenamefont {Thompson},\ and\ \citenamefont {Sarenac}}]{pushin2023structured}%
  \BibitemOpen
  \bibfield  {author} {\bibinfo {author} {\bibfnamefont {Dmitry~A}\ \bibnamefont {Pushin}}, \bibinfo {author} {\bibfnamefont {David~G}\ \bibnamefont {Cory}}, \bibinfo {author} {\bibfnamefont {Connor}\ \bibnamefont {Kapahi}}, \bibinfo {author} {\bibfnamefont {Mukhit}\ \bibnamefont {Kulmaganbetov}}, \bibinfo {author} {\bibfnamefont {Melanie}\ \bibnamefont {Mungalsingh}}, \bibinfo {author} {\bibfnamefont {Andrew~E}\ \bibnamefont {Silva}}, \bibinfo {author} {\bibfnamefont {Taranjit}\ \bibnamefont {Singh}}, \bibinfo {author} {\bibfnamefont {Benjamin}\ \bibnamefont {Thompson}}, \ and\ \bibinfo {author} {\bibfnamefont {Dusan}\ \bibnamefont {Sarenac}},\ }\bibfield  {title} {\enquote {\bibinfo {title} {Structured light enhanced entoptic stimuli for vision science applications},}\ }\href@noop {} {\bibfield  {journal} {\bibinfo  {journal} {Frontiers in Neuroscience}\ }\textbf {\bibinfo {volume} {17}} (\bibinfo {year} {2023})}\BibitemShut {NoStop}%
\bibitem [{\citenamefont {Pushin}\ \emph {et~al.}(2025)\citenamefont {Pushin}, \citenamefont {Garrad}, \citenamefont {Kapahi}, \citenamefont {Silva}, \citenamefont {Chahal}, \citenamefont {Cory}, \citenamefont {Kulmaganbetov}, \citenamefont {Salehi}, \citenamefont {Mungalsingh}, \citenamefont {Singh} \emph {et~al.}}]{pushin2025characterizing}%
  \BibitemOpen
  \bibfield  {author} {\bibinfo {author} {\bibfnamefont {Dmitry~A}\ \bibnamefont {Pushin}}, \bibinfo {author} {\bibfnamefont {Davis~V}\ \bibnamefont {Garrad}}, \bibinfo {author} {\bibfnamefont {Connor}\ \bibnamefont {Kapahi}}, \bibinfo {author} {\bibfnamefont {Andrew~E}\ \bibnamefont {Silva}}, \bibinfo {author} {\bibfnamefont {Pinki}\ \bibnamefont {Chahal}}, \bibinfo {author} {\bibfnamefont {David~G}\ \bibnamefont {Cory}}, \bibinfo {author} {\bibfnamefont {Mukhit}\ \bibnamefont {Kulmaganbetov}}, \bibinfo {author} {\bibfnamefont {Iman}\ \bibnamefont {Salehi}}, \bibinfo {author} {\bibfnamefont {Melanie~A}\ \bibnamefont {Mungalsingh}}, \bibinfo {author} {\bibfnamefont {Taranjit}\ \bibnamefont {Singh}},  \emph {et~al.},\ }\bibfield  {title} {\enquote {\bibinfo {title} {Characterizing the circularly oriented macular pigment using spatiotemporal sensitivity to structured light entoptic phenomena},}\ }\href@noop {} {\bibfield  {journal} {\bibinfo  {journal} {Journal of Vision}\ }\textbf {\bibinfo {volume} {25}},\
  \bibinfo {pages} {11--11} (\bibinfo {year} {2025})}\BibitemShut {NoStop}%
\bibitem [{\citenamefont {Bone}(1980)}]{bone1980role}%
  \BibitemOpen
  \bibfield  {author} {\bibinfo {author} {\bibfnamefont {Richard~A}\ \bibnamefont {Bone}},\ }\bibfield  {title} {\enquote {\bibinfo {title} {The role of the macular pigment in the detection of polarized light},}\ }\href@noop {} {\bibfield  {journal} {\bibinfo  {journal} {Vision Research}\ }\textbf {\bibinfo {volume} {20}},\ \bibinfo {pages} {213--220} (\bibinfo {year} {1980})}\BibitemShut {NoStop}%
\end{thebibliography}%
\clearpage
\onecolumngrid
\appendix

\section*{Appendix}

\subsection{Experimental Protocol}

The entire experiment was controlled by a custom Python program running on Windows, which handled all stages of the procedure including loading calibration data, implementing staircase logic, and automatically saving participant output to ID-labeled folders. All optomechanical components were sourced from a single manufacturer (Thorlabs Inc.) to ensure full integration and compatibility, and the Pylablib package was used to interface with Thorlabs devices through Python.

For each participant, testing began with the selection of an annular aperture according to a predefined, counterbalanced order covering six conditions. Once the aperture was set, both light paths were activated to generate a rotating spin–orbit stimulus at 80\,\% contrast. This high-contrast rotating pattern served as a brief preparation phase prior to the first trial of each aperture condition.  

After the participant confirmed proper head alignment with the optical axis, the test was initiated by pressing the central key on a three-button keypad. This triggered closure of the mechanical shutter blocking the polarized light path. The program then randomly selected one of two interleaved staircases: Staircase A, beginning at 80\,\% contrast, or Staircase B, beginning at 30\,\%. For each subsequent trial, contrast levels were adjusted dynamically based on the history of the selected staircase.

Once the contrast was set by modulating the current to the two light sources, the motor accelerated to the target rotation speed of 540$^\circ$/s. The shutter then opened for a 250\,ms exposure and closed immediately afterward, halting the rotation. Participants indicated the perceived rotation direction using the keypad (right = clockwise, left = counterclockwise).

The program evaluated the response, logged its correctness, updated the reversal status, and recorded all relevant data, including staircase history. After each trial, termination criteria were checked; if met, the current staircase was concluded. The program then selected the next staircase at random and repeated the process. Upon completion of both staircases for a given aperture, the system automatically advanced to the next aperture and continued until all six conditions were completed.

\begin{figure}[!h]
    \centering
    \includegraphics[width=.9\linewidth]{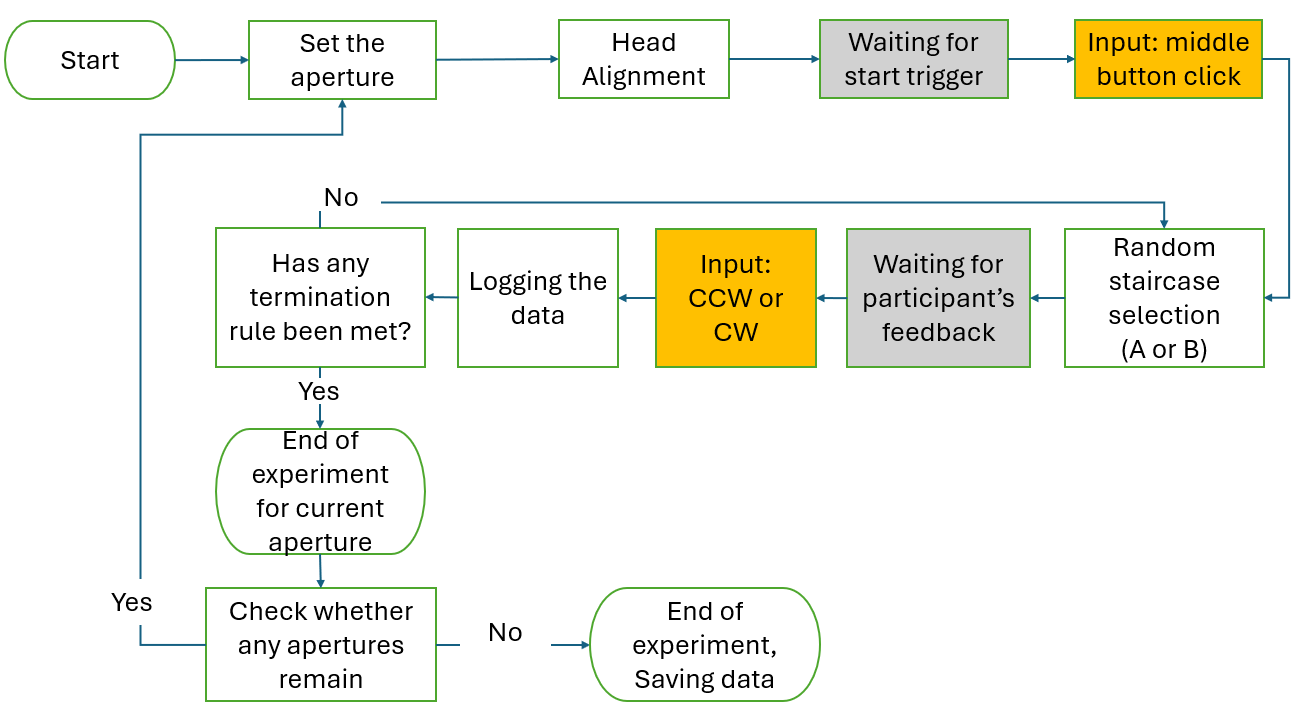}
     \caption{Flowchart of the fully automated experimental procedure. A Python program controlled all aspects of the task, including loading calibration data, setting aperture order, managing the rotating spin–orbit stimulus, implementing the interleaved staircases, and logging responses. Participants initiated each trial and provided feedback via a three‑button keypad (right button = clockwise, left button = counterclockwise), while all other steps—including contrast adjustment, stimulus presentation, and progression through apertures—were executed automatically by the system.}

    \label{fig:flowchart}
\end{figure}

\end{document}